\documentclass[11pt]{article}
\usepackage{amsmath,amssymb,color,epsfig,cite}
\usepackage{graphicx}
\usepackage{subfigure}
\usepackage{setspace}

\usepackage{amsfonts}
\newcommand{\hoch}[1]{$\, ^{#1}$}

\makeatletter
\@addtoreset{equation}{section}
\makeatother

\newcommand{\be}{\begin{equation}}
\newcommand{\ee}{\end{equation}}
\newcommand{\bea}{\setlength\arraycolsep{2pt} \begin{eqnarray}}
\newcommand{\eea}{\end{eqnarray}}
\newcommand{\nn}{\nonumber}

\def\ft#1#2{{\textstyle{\frac{\scriptstyle #1}{\scriptstyle #2} } }}
\def\fft#1#2{{\frac{#1}{#2}}}

\def\0{{\sst{(0)}}}
\def\1{{\sst{(1)}}}
\def\2{{\sst{(2)}}}
\def\3{{\sst{(3)}}}
\def\4{{\sst{(4)}}}
\def\5{{\sst{(5)}}}
\def\6{{\sst{(6)}}}
\def\7{{\sst{(7)}}}
\def\8{{\sst{(8)}}}
\def\sst#1{{\scriptscriptstyle #1}}

\def\del{{\partial}}

\begin{document}

\begin{flushright}
\hfill{MI-TH-1761}

\end{flushright}

\vspace{25pt}
\begin{center}
{\Large {\bf Aretakis Charges and Asymptotic Null Infinity}}

\vspace{15pt}
{\bf Hadi Godazgar\hoch{1}, Mahdi Godazgar\hoch{2} and 
C.N. Pope\hoch{3,4}}

\vspace{10pt}

\hoch{1} {\it Max-Planck-Institut f\"ur Gravitationsphysik (Albert-Einstein-Institut), \\
M\"ühlenberg 1, D-14476 Potsdam, Germany. \\
hadi.godazgar@aei.mpg.de}

\vspace{10pt}

\hoch{2} {\it Institut f\"ur Theoretische Physik,\\
Eidgen\"ossische Technische Hochschule Z\"urich, \\
Wolfgang-Pauli-Strasse 27, 8093 Z\"urich, Switzerland.\\
godazgar@phys.ethz.ch}

\vspace{10pt}

\hoch{3} {\it George P. \& Cynthia Woods Mitchell  Institute
for Fundamental Physics and Astronomy,\\
Texas A\&M University, College Station, TX 77843, USA.\\
pope@physics.tamu.edu}

\vspace{10pt}

\hoch{4}{\it DAMTP, Centre for Mathematical Sciences,\\
 Cambridge University, Wilberforce Road, Cambridge CB3 OWA, UK}

 \vspace{20pt}

\vspace{20pt}

\underline{ABSTRACT}
\end{center}
We construct a relation between the Aretakis charge of any extreme black hole 
and the Newman-Penrose charge.  This is achieved by constructing a conformal 
correspondence between extreme black holes and what we call weakly 
asymptotically flat space-times.  Under this correspondence the Newman-Penrose 
charge of the weakly asymptotically flat space-time maps on to the 
Aretakis charge 
of the extreme black hole.  Furthermore, we generalise the conformal isometry 
displayed by the extreme Reissner-Nordstr\"om solution to a novel conformal 
symmetry that acts within a class of static STU supergravity black holes.

\thispagestyle{empty}

\pagebreak

\tableofcontents
\addtocontents{toc}{\protect\setcounter{tocdepth}{2}}



\section{Introduction}

The horizon of an extreme black hole is unstable under scalar perturbations 
\cite{Aretakis:RN1, Aretakis:RN2, Aretakis:extremal, Aretakis:far}.  More 
generally, at least for the case of extreme Reissner-Nordstr\"om (ERN) and 
extreme Kerr black holes, the horizon is also unstable under massive scalar, 
electromagnetic and linearised gravitational perturbations 
\cite{LR:grav, LuciettiERN, Murata}.  This instability is due to the existence 
of conserved charge(s) on the extremal horizon---the \emph{Aretakis charge(s)}.

Beyond the calculational intricacies, our aim in this paper is to understand 
why such conserved charges exist on the extremal horizon; say as opposed to 
bifurcate horizons, where Aretakis charges do not exist.  If we consider, 
for example, the ERN or extreme Kerr black holes, apart from the associated Aretakis 
charges on the horizon there exist another set of conserved charges, 
namely 
the \emph{Newman-Penrose} (NP) charges \cite{NP} which are defined at 
future null 
infinity, which is strictly not a part of these space-times.  In fact, for any 
asymptotically flat space-time (see e.g.\ \cite{Wald}), there 
exist NP charge(s) 
at its future null infinity.

Unlike the  Aretakis charges, the existence of NP charges is not so 
surprising.  In fact we would expect such charges to exist from a 
physical point of view, given that asymptotically flat space-times 
are defined 
precisely such that there is a notion of approaching ``flat space-time'' at 
large radial distance away from the gravitational source of interest.  Since 
conserved charges exist in flat space-time, 
it is reasonable to expect that they 
will exist at future null infinity \cite{WZ} (see also \cite{goldberg}).

For scalar perturbations on the ERN background, it has been shown 
\cite{BF, LuciettiERN} that there is a simple bijective map relating the 
Aretakis and NP charges.  The existence of such a map is due to the fact 
that the ERN metric displays a conformal isometry, generated by a 
spatial inversion that interchanges the extremal horizon and future null 
infinity \cite{Couch1984}.  The massless scalar wave equation is 
conformally covariant on a Ricci scalar-flat background.  Therefore, as 
future null infinity maps on to the horizon under the spatial inversion, 
so do its NP charges map on to the Aretakis charges of the horizon.  

The above argument provides an attractive explanation for the existence of 
Aretakis charges.  However, the conformal isometry at the heart of this 
explanation is a rather special feature of the ERN black hole.  Such an 
isometry does not exist in more general situations.  For example, the 
extreme Kerr black hole does not have such a conformal isometry 
\cite{Couch1984}.~\footnote{Although, an axially symmetric scalar field 
propagating on the extreme Kerr(-Newman) background does display such a 
symmetry \cite{BF}. This is closely related to the fact that the radial 
equation displays such a property \cite{Couch1984}.}  Furthermore, 
the argument for the existence of Aretakis charges associated with a 
massless scalar field for general extreme black holes 
\cite{LR:grav, Aretakis:extremal} does not preclude non-asymptotically 
flat extreme black holes.  Such space-times have no NP charges.  Hence, 
an argument of the form above, relating different charges in the 
space-time to 
one another, is impossible here.

In this paper, we aim to study the relation between Aretakis and NP charges 
in a more general context.  Throughout the paper, for lucidity, we 
confine ourselves to four dimensions and massless scalar perturbations.  
These restrictions do not affect our main results, which will follow 
through, almost trivially, for higher-dimensional space-times, as well as to 
other conformally-covariant fields.  

We find that a more general class of solutions, the 4-charge static 
extreme STU black holes, also exhibit a kind of conformal symmetry of the 
type exhibited by the ERN black hole, albeit of a more general type.  Under a 
spatial inversion, analogous to that performed in \cite{Couch1984} for the 
ERN black hole, we find that a general 4-charge STU black hole maps on to a 
space-time 
that is conformally related to another such STU black hole but 
with different values for the charge parameters.  Thus, while the 
transformation does not amount to an isometry, there is an intriguing 
symmetry within the class of static 4-charge 
STU black holes which means that the 
Aretakis charges of one black hole are related to the NP charges of another, 
and \textit{vice-versa}.  For a subclass of the 4-charge black holes, where 
the charges are set pairwise-equal (which includes the ERN black hole
as a special case), 
we find that there is indeed a symmetry enhancement to a conformal isometry.  

Motivated by the above observation for STU black holes we relate, 
via a spatial inversion, any extreme black hole space-time and its Aretakis 
charge(s), to a \emph{weakly asymptotically flat} space-time and its associated 
NP charge(s).~\footnote{Note that our definition of weakly asymptotic
has no relation to the usage 
in \cite{ch,bizon}.}  If one  makes the fall-off conditions for certain 
metric components slightly stronger, weakly asymptotically flat 
space-times 
reduce to asymptotically flat space-times in Bondi coordinates.

In section \ref{sec:charge}, we review the construction of the NP and 
Aretakis charges.  In section \ref{sec:NP}, we derive the NP charge 
for \emph{weakly asymptotically flat space-times}.  Compared to 
asymptotically flat space-times, the fall-off conditions that we assume for 
the metric components are more relaxed.  Moreover, the spatial sections 
at null infinity are merely assumed to be compact rather than 
specifically $S^2.$ This result emphasises the fact that NP charges do not 
only exist for asymptotically flat space-times, but also for more general 
space-times.  
Furthermore, we shall use this result in section \ref{sec:corres}.  
In section \ref{sec:sphsym}, we show that for stationary spherically-symmetric 
space-times one has a hierarchy of NP charges.  This result complements a 
similar result by Aretakis regarding stationary spherically symmetric 
extreme black holes \cite{Aretakis:extremal}.  In section \ref{sec:A}, we 
review the construction in \cite{LR:grav} of the Aretakis charge using 
Gaussian null coordinates introduced in the vicinity of the horizon 
\cite{Kunduri:GNC}.  We rederive the first Aretakis charge of the ERN black 
hole as an example.  

In section \ref{sec:stu}, we consider a class of static (extreme) 
black hole solutions of STU supergravity, namely a four-charge family 
that includes as a sub-class the one-parameter ERN family.  We find a novel 
conformal symmetry under spatial inversion, whereby the horizon of a 
particular solution with given parameters maps on to the future null 
infinity of another solution with related, but different, parameters, 
and \textit{vice-versa}.  This conformal symmetry that acts within the 
family of static 4-charge STU black holes can be thought of as 
a generalisation of 
the conformal isometry displayed by the ERN solution.  In fact, if we 
specialise to a two-parameter subclass, by setting the four charges
pairwise equal, we find that the black hole parameters map onto 
themselves under the spatial inversion, and hence the conformal 
symmetry enhances to a conformal isometry.  The ERN black holes are 
contained as a special case within the pairwise-equal class, in which
all the charges are set equal.  In contrast to the four-charge solutions, 
we find that no such generalised conformal symmetry exists for the most 
general eight-charge dyonic static STU black holes, the details of which
are given in appendix \ref{app:dyonic}.

In section \ref{sec:corres}, we derive the main result of the paper: we 
find a conformal correspondence between weakly asymptotically flat 
space-times 
and extreme black holes.  First, we write the extreme black hole metric in 
Gaussian null coordinates, which can always be chosen in the neighbourhood 
of the horizon.  In these coordinates, the horizon corresponds to the 
null surface $r=0$.  We then perform a coordinate transformation, which 
includes the spatial inversion $r \rightarrow 1/r$.  We identify the 
resulting metric as being conformally related to a space-time whose metric is 
naturally written in Bondi coordinates---a weakly asymptotically flat 
space-time.  Since the original metric was valid in the neighbourhood of 
$\{r=0\}$, we expect the weakly asymptotically flat metric to describe 
a space-time for which we can choose asymptotic coordinates (i.e.~coordinates 
valid for large $r$), such that the metric coincides with that of the weakly 
asymptotically flat metric for large $r$.  Moreover, note that the weakly 
asymptotically flat space-time we obtain via this map will not be Ricci-flat.  
However, it will be asymptotically Ricci-flat.

The results of section \ref{sec:NP} guarantee the existence of NP charges 
for weakly asymptotically flat space-times.  Thus, this correspondence maps the 
NP charge(s) of the associated weakly asymptotically flat space-time to 
the Aretakis charge(s) of the original extreme black hole under 
consideration.  In section \ref{sec:scalar}, we make this map between the 
Aretakis and NP charges precise for the massless scalar, and we consider the 
ERN black hole as an example.  It should be stressed that the associated 
weakly asymptotically flat space-time has no direct physical significance.  The 
construction merely highlights the correspondence between Aretakis and 
NP charges. 

We end with some discussions in section \ref{sec:dis}.

\section{Conserved charges at null infinity and the extremal horizon} 
\label{sec:charge}

\subsection{Newman-Penrose charge} \label{sec:NP}

In this section, we derive the well-known Newman-Penrose (NP) 
charge \cite{NP} for a massless scalar field at future null infinity.  
The existence of a boundary that can be identified with future null 
infinity necessitates that the solution be asymptotically flat.  Thus, 
NP charges are generally defined for such space-times. However here we 
slightly weaken the definition of asymptotic flatness, and show that NP charges 
can nevertheless still be defined.  The main reason for doing this is 
that later we will need a notion of an NP charge for such slightly more 
general space-times.  Another motivation for this exercise is to highlight 
 the fact that NP charges arise in this more general context, and they are
not merely tied to the notion of asymptotic flatness.

We define a \emph{weakly asymptotically flat} space-time as one for which
Bondi coordinates \cite{bondi, sachs} $(u,r,x^I)$ can be introduced, 
such that the metric takes the Bondi form
\begin{equation} \label{AF}
 d s^2 = - F e^{2 \beta} du^2 - 2 e^{2 \beta} du dr + 
r^2 h_{IJ} \, (dx^I - C^I du) (dx^J - C^J du),
\end{equation}
with the metric functions satisfying the fall-off conditions
\begin{gather}
 \lim_{r\to\infty} r(F-1) = \tilde{F}(u,x^I), \qquad \lim_{r\to\infty} r^2 \beta = \tilde{\beta}(u,x^I), \notag \\[2mm] \label{falloff}
 \lim_{r\to\infty} r C^I = \tilde{C}^I(u,x^I), \qquad \lim_{r\to\infty} r(h_{IJ}-\omega_{IJ}) = \tilde{h}_{IJ}(u,x^I),
\end{gather}
at large $r$, 
where the 2-dimensional space has the  metric $\omega_{IJ}$ with 
coordinates $x^I$, and is assumed to be compact.  

The differences between weakly asymptotically flat and asymptotically flat 
space-times are two-fold.  First, for asymptotically flat space-times, 
$\omega_{IJ}$ is the metric on 
the unit round 2-sphere.  Here, we assume only
that $\omega_{IJ}$ is a positive-definite metric on a complete 
compact 2-manifold.   Moreover, our assumption about the fall-off for 
$C^{I}$ at large $r$ is weaker than that for asymptotically flat space-times, 
for which it is required that
\begin{equation}
 \lim_{r\to\infty} r^2 C^I = \tilde{C}^I(u,x^I).
\end{equation}

We can always choose a gauge in which
\begin{equation} \label{det:h}
 h \equiv \textup{det}(h_{IJ}) = \omega\, \zeta(r)^2,
\end{equation}
where $\omega=\textup{det}(\omega_{IJ})$ and $\zeta(r)$ is some function of $r$ such that
\begin{equation}
 \lim_{r\to\infty} r(\zeta-1) = \textup{constant}.
\end{equation}
Note that, we could of course, eliminate $\zeta$ by redefining $r$. 
However, for later convenience we choose to fix our gauge as in 
equation \eqref{det:h}.

We now show that this more general class of space-times, which includes 
asymptotically flat space-times as a strict subset, contains an NP charge 
for a massless scalar field at future null infinity, which we define to be 
the surface $r=c, \ c \rightarrow \infty$.

Consider a massless scalar on the above background, satisfying
\begin{equation} \label{KG:NP}
 \Box_g \psi = \frac{1}{\sqrt{-g}} \partial_a \left( \sqrt{-g}\, g^{ab} \, 
\partial_b \psi \right) = 0.
\end{equation}
We assume the fall-off condition
\begin{equation}
 \lim_{r\to\infty} (r \psi) = \psi^{(1)}(u,x^I),
\end{equation}
at large $r$.  Using the fact that, with respect to coordiantes $(u,r,x^I)$,
\begin{equation}
 (e^{2\beta} g^{ab}) = \begin{pmatrix}
               0 & -1 & 0 \\
              -1 & F & - C^J \\
               0 & - C^I & e^{2\beta} h^{IJ}/r^2
              \end{pmatrix}
\end{equation}
and that
\begin{equation} \label{det:asymp}
 \sqrt{-g} = r^2\, \zeta(r)\, e^{2\beta} \sqrt{\omega},
\end{equation}
equation \eqref{KG:NP} reduces, at leading order as $r\rightarrow\infty$, to
\begin{equation} \label{eq:wave}
 - \partial_u \big[ \partial_r(r^2 \zeta)\, \psi + 2r^2 \zeta\, 
\partial_{r} \psi \big] -\frac{r^2\, \zeta}{\sqrt{\omega}} \partial_I 
\big[ \sqrt{\omega} \, C^{I} \partial_{r} \psi \big] + 
\frac{\zeta}{\sqrt{\omega}} \partial_I \big[ e^{2 \beta} 
\sqrt{\omega} h^{IJ} \partial_{J} \psi \big] = 0.
\end{equation}
Note that the second term, which involves $C^I$, would drop out for asymptotically flat space-times.  Multiplying equation \eqref{eq:wave} by $r$ and taking the limit $r \rightarrow \infty$ gives
\begin{equation}
 \lim_{r\to\infty} \left\{ - r^2 \partial_u  \left[2\, \partial_r (r \psi ) + r \partial_r \zeta\, \psi \right]
 + \frac{r}{\sqrt{\omega}} \partial_I \big[ \sqrt{\omega} \left( \omega^{IJ} \partial_{J} \psi - r^2 C^I \partial_r \psi \right) \big] \right\} = 0.
\end{equation}
Now, integrating this equation over the compact space at infinity, 
whose metric is $\omega_{IJ}$, we find that the second set of terms, which do not involve a $u$-derivative, reduce to a 
boundary contribution, which can be discarded.  Thus, we have a conserved 
charge at future null infinity that is given by
\begin{equation} \label{NPcharge}
 H_{NP} = - \lim_{r\to\infty} \int d\omega \ r^2 \left[2 \partial_r (r \psi ) + r \partial_r \zeta\, \psi \right],
\end{equation}
where $d \omega = \sqrt{\omega}\, dx^I$.

As mentioned earlier, we can always choose a gauge in which $\zeta=1.$  
In this gauge, the NP charge reduces to
\begin{equation}
 H_{NP}|_{\zeta=1} = - \lim_{r\to\infty} \int d\omega \ 2 r^2 \partial_r (r \psi ).
\end{equation}
Note that this charge does not depend on the specific features of the 
space-time and it is therefore universal.

If we assume that the scalar field is analytic in $1/r$, and so it 
admits an expansion 
\begin{equation}
 \psi = \sum_{n=1}^{\infty} \frac{\psi^{(n)}(u,x^I)}{r^n},
\end{equation}
the Newman-Penrose charge reduces to
\begin{equation}
 H_{NP} = 2 \int_{S^2} d\omega \, \psi^{(2)}.
\end{equation}

\subsubsection{Example: Spherically symmetric case} \label{sec:sphsym}

For the case of spherically symmetric backgrounds, one has a hierarchy of NP 
charges.  A similar result also holds for Aretakis charges 
\cite{Aretakis:extremal}.  Given a stationary spherically symmetric 
space-time, one can choose coordinates $\{u,r,\theta,\phi\}$ such 
that the metric takes the form
\begin{equation}
ds^2 = - F(r)\, du^2 - 2 du dr + r^2 \zeta(r)\, d \Omega^2,
\end{equation}
where 
\begin{equation}
 \lim_{r\to\infty} r (F-1) = \tilde{F}^{(1)}, \qquad  \lim_{r\to\infty} r (\zeta-1) = \tilde{\zeta}^{(1)}.
\end{equation}
The scalar wave equation \eqref{KG:NP} on this background reduces to
\begin{equation} \label{wave:sph}
 - \partial_u \Big[ 2 r^2 \zeta\, \partial_r \psi + \partial_{r}(r^2 \zeta)\, \psi \Big] + \partial_r \big( r^2 \zeta\, F\, \partial_r \psi \big) + \Delta \psi = 0,
\end{equation}
where $\Delta$ is the scalar Laplacian on the unit 2-sphere.

In order to proceed further, we assume an analytic expansion for the 
scalar field,
\begin{equation} \label{psi:exp}
 \psi = \sum_{n=1}^{\infty} \frac{\psi^{(n)}(u,\theta,\phi)}{r^n},
\end{equation}
as well as for the metric functions $F$ and $\zeta$.  We emphasise that this 
assumption is not essential (see \cite{Aretakis:extremal}).  However, we do 
so to make the appearance of the hierarchy of NP charges clearer.

Substituting the expansion \eqref{psi:exp} into the wave 
equation \eqref{wave:sph} gives
\begin{align}
 \sum_{n=1}^{\infty} \frac{1}{r^n} \left\{ \partial_u \big[ 2 n 
\zeta\, \psi^{(n+1)} - (r^2 \partial_r \zeta)\, \psi^{(n)} \big] 
 + n(n-1)\, \zeta F\, \psi^{(n)}  + 
\Delta \psi^{(n)} \right.& \notag \\
 &\hspace{-30mm} \left. - (n-1) \big[ r^2 
\partial_r(\zeta F) \big] \psi^{(n-1)} \right\} = 0.
\end{align}
 From this, we can extract a hierarchy of equations labelled by a 
positive integer $n$:
\begin{equation}
T_{n} = 0,
\end{equation}
where
\begin{equation}
 T_{n} = \partial_{u} \left( \sum_{i=1}^{n+1} \alpha_n^{(i)}\, 
\psi^{(i)} \right) + \big[ n(n-1) +
 \Delta \big] \psi^{(n)} + \sum_{i=1}^{n-1} \beta_n^{(i)} \, \psi^{(i)},
\end{equation}
with $\alpha_n^{(i)}$ and $\beta_n^{(i)}$ being certain functions of the 
metric components, which we leave implicit.

For $n=1$, we have
\begin{equation}
 2 \partial_{u} \left( \psi^{(2)} + 
\lim_{r\rightarrow \infty}(r^2 \partial_r \zeta)\, \psi^{(1)} \right) + 
\Delta \psi^{(1)} = 0.
\end{equation}
Integrating this equation over the sphere, we obtain the NP charge 
\eqref{NPcharge} that we derived more generally in the previous analysis.

For general $n>1$, multiplying by the spherical harmonics 
$Y_{\ell, m}$ with $\ell=n-1$ and integrating over the sphere gives
\begin{equation} \label{Tn:eqn}
 \int d \omega\ Y_{n-1,m} \left\{ \partial_{u} 
\left( \sum_{i=1}^{n+1} \alpha_n^{(i)}\, \psi^{(i)} \right) + 
\sum_{i=1}^{n-1} \beta_n^{(i)} \, \psi^{(i)} \right\} = 0, 
\end{equation}
where we have used the fact that
\begin{equation}
\int d \omega \, Y_{\ell,m}\ \Delta f = -\ell (\ell + 1) \int d \omega \,  Y_{\ell,m}\, f
\end{equation}
for any function $f$.  Hence, the only obstacle to obtaining an 
extra $(2n-1)$ NP charges at each level $n$ is the second set of terms.  
However, these terms can be iteratively rewritten to be of the same form as 
the first set of terms in \eqref{Tn:eqn}, by making use of 
the lower-order equations $T_{p}=0, \ p<n$.  For example, we show how one 
can deal with the $\beta_n^{(n-1)} \psi^{(n-1)}$ term in equation 
\eqref{Tn:eqn}.  Multiplying equation $T_{n-1}=0$ by $Y_{n-1,m}$ and 
integrating over the 2-sphere gives
\begin{equation}
  \int d \omega\ Y_{n-1,m} \left\{  \partial_{u} 
\left( \sum_{i=1}^{n} \alpha_{n-1}^{(i)}\, \psi^{(i)} \right) + 
\sum_{i=1}^{n-1} \beta_{n-1}^{(i)} \, \psi^{(i)} \right\} = 0,
\end{equation}
where $\beta_{n-1}^{(n-1)}= -2(n-1) \neq 0.$  Therefore, this equation 
allows us to rewrite equation \eqref{Tn:eqn} as
\begin{equation}
 \int d \omega\  Y_{n-1,m} \left\{ \partial_{u} \left( 
\sum_{i=1}^{n+1} \tilde{\alpha}_n^{(i)}\, \psi^{(i)} \right) + 
\sum_{i=1}^{n-2} \tilde{\beta}_n^{(i)} \, \psi^{(i)} \right\} = 0 
\end{equation}
with $\tilde{\alpha}_n^{(i)},\, \tilde{\beta}_n^{(i)}$ some functions of the metric components.  Moreover, using equation $T_{n-2}=0$, we can rewrite the $i=(n-2)$ term 
in the second set of terms above in terms of lower-order terms and 
other terms of the form $\partial_{u}(\ldots).$  Following this 
prescription down to equation $T_1=0$ gives a set of $(2n-1)$ NP 
charges for each $n$:
\begin{equation}
 \partial_u\, c_{n,m}  = 0  
\end{equation}
where
\begin{equation}
 c_{n,m}=\int d \omega\  Y_{n-1,m}\ \sum_{i=1}^{n+1} a_n^{(i)}\, \psi^{(i)}
\end{equation}
for certain constants $a_n^{(i)}$ that are determined by implementing the
above iterative procedure. Note that
\begin{equation}
 a_n^{(n+1)} = 2n.
\end{equation}
Therefore, unless $\psi^{(n+1)} = 0$ and $a_n^{(i)}\, \psi^{(i)} = 0$ for 
all $i<(n+1)$, which would not occur in a generic case, we have 
\begin{equation}
 c_{n,m} \neq 0.
\end{equation}

\subsection{Aretakis charge} \label{sec:A}

In this section, we rederive the result \cite{LR:grav, Aretakis:extremal} 
that all extreme black holes admit a conserved quantity, associated with a 
massless scalar test field, on the horizon.  Assuming that the horizon 
is Killing, one can introduce Gaussian null coordinates $(v,r,x^I)$ in 
the vicinity of the horizon (the surface $r=0$), such that the metric takes 
the form \cite{Kunduri:GNC}
\begin{equation} \label{NHC}
 ds^2 = L(x)^2 \Big[- r^2 F(r,x) dv^2 + 2 dv dr \Big]  + 
\gamma_{IJ}(r,x) \Big(dx^I - r\, h^I(r,x) dv \Big) \Big(dx^J - 
r\, h^J(r,x) dv\Big),
\end{equation}
with $F(r=0,x^I) = 1$.  The timelike Killing vector field is
\begin{equation}
 k = \frac{\partial}{\partial v}
\end{equation}
in these coordinates.  Moreover, we are guaranteed the existence of a 
rotational Killing vector field \cite{hawking} 
(see also \cite{Alexakis:2013rya}), which we take to be
\begin{equation}
 m = \frac{\partial}{\partial \phi}
\end{equation}
with $\phi \equiv x^1$.  Furthermore, we assume that~\footnote{In fact, we can assume the slightly weaker condition that $\lim_{r\to 0} h^{2} = 0.$}
\begin{equation} \label{h2falloff}
  \lim_{r \to 0} r^{-1} h^{2} = \tilde{h}^2(x^2).
\end{equation}

   Now consider a massless scalar field $\psi$ on this background, well-behaved on the horizon, obeying
\begin{equation} \label{scalareqn}
 \Box_g \psi = \frac{1}{L^2 \sqrt{\gamma}} \partial_a 
\left( L^2 \sqrt{\gamma}\, g^{ab} \, \partial_b \psi \right) = 0,
\end{equation}
where $\gamma$ is the determinant of the 2-dimensional metric $\gamma_{IJ}$.
Using the fact that
\begin{equation}
 (L^2 g^{ab}) = \begin{pmatrix}
               0 & 1 & 0 \\
               1 & r^2 F & r h^J \\
               0 & r h^I & L^2 \gamma^{IJ}
              \end{pmatrix},
\end{equation}
where $\gamma^{IJ}$ is the inverse of $\gamma_{IJ}$, equation \eqref{scalareqn} reduces \emph{on the horizon} $r=0$ to
\begin{equation}
 \frac{1}{\sqrt{\gamma}} \left\{ \partial_v \big[ 2 \sqrt{\gamma} 
\partial_r \psi + \partial_r \sqrt{\gamma} \psi \big] + 
\partial_\phi (\sqrt{\gamma}\, h^{\phi}\, \psi) 
 + \partial_I \big[ \sqrt{\gamma} L^2 \gamma^{IJ} \partial_J \psi 
\big] \right\} = 0,
\end{equation}
where we have used the fact that $k$ and $m$ are Killing vectors, and so 
the metric components do not depend on $v$ and $\phi$, as well as the 
fall-off condition \eqref{h2falloff}.  Integrating this equation over 
coordinates $x^I$ with measure $\sqrt{\gamma}$, we identify the second and 
third set of terms, which do not involve $v$-derivatives, as total derivative terms.  Since the 
horizon is compact we can disregard these terms, which leaves us with a 
conserved quantity, the Aretakis charge, on the horizon:
\begin{equation} \label{Acharge}
 H_{Aretakis} = \lim_{r \rightarrow 0} \int dx^2  \sqrt{\gamma}\ \big[ 2 \, \partial_r \psi + 
\textstyle{\frac{1}{2}}\, \partial_r \log\gamma\, \psi \big].
\end{equation}

\subsubsection{Example: Extreme Reissner-Nordstr\"om solution}

The ERN metric can be put into the form \eqref{NHC} by 
redefining $r$ in the standard metric
\begin{equation} \label{ERN}
 ds^2 = - \frac{(r-M)^2}{r^2} dv^2 + 2 dv dr + r^2 d \Omega^2,
\end{equation}
where $d\Omega^2 = d\theta^2 + \sin ^2 \theta\, d \phi^2$, as follows
\begin{equation}
 r \rightarrow M + M^2 r,
\end{equation}
so that the horizon is now at $r=0$.\footnote{Note that the new radial
coordinate has dimensions of inverse length. This is done in order to be
consistent with the assumed form of the metric in \eqref{NHC}, where
the function $F(r,x)$ is dimensionless.}  This gives
\begin{equation} \label{ERN:GNC}
 ds^2 = M^2 \left[ - \frac{r^2}{(1+Mr)^2} dv^2 + 2 dv dr \right] + M^2 (1+Mr)^2 d \Omega^2.
\end{equation}
Comparing with metric \eqref{NHC}, we have
\begin{equation}
 L=M, \quad F = (1+Mr)^{-2}, \quad h^I = 0, \quad 
\gamma = M^4 (1+Mr)^{4} \sin^2 \theta.
\end{equation}
It is now simple to show, using equation \eqref{Acharge}, that the Aretakis 
charge for the ERN black hole \cite{Aretakis:RN1, Aretakis:RN2} 
is~\footnote{The form of the Aretakis charge in 
Refs.~\cite{Aretakis:RN1, Aretakis:RN2, Aretakis:extremal} is slightly 
different from that here. This is simply due to the different coordinate 
systems used.  Here, we have shifted the horizon to $r=0.$}
\begin{equation} \label{AERN}
 H_{Aretakis} = 2 M^2\, \lim_{r \rightarrow 0} \int_{S^2} d \omega\, (\partial_r \psi + M \psi).
\end{equation}

\section{Extreme black holes in STU supergravity} \label{sec:stu}

The existence of conserved charges was an unexpected property of extremal 
horizons. For the case of the ERN solution, at least, the existence of 
Aretakis charges has been successfully understand \cite{BF,LuciettiERN}, 
in terms of NP charges, using the Couch-Torrence conformal 
isometry \cite{Couch1984} that interchanges the horizon, on which Aretakis 
charges are defined, and future null infinity, on which NP charges are 
defined.  Unfortunately, such a conformal isometry does not exist for general 
extreme black holes, such as the extreme Kerr black hole.  We shall deal 
with the general case in the next section.  For now, in this section, we 
present a slightly more involved Couch-Torrence-like conformal relation that 
takes a particular solution within a family of extreme solutions to 
a different solution within the family.  For a sub-family 
of solutions among these examples, the conformal relation 
in fact enhances to an isometry.

\subsection{The static 4-charge black holes}

Here, we shall consider 4-charge static black holes in ungauged 
four-dimensional
STU supergravity.  The relevant part of the bosonic Lagrangian is
\be
{\cal L}= \sqrt{-g}\, \Big[R - \ft12 (\del\vec\varphi)^2 -
  \ft14 \sum_{i=1}^4 e^{\vec a_i\cdot\vec\varphi}\, 
F^{i\, \mu\nu} F^i_{\mu\nu}\Big]\,,
\ee
where the vectors $\vec a_i$ characterising the coupling of the three
dilatons $\vec\varphi$ to the four gauge fields obey the relation
\be
\vec a_i\cdot \vec a_j = 4\delta_{ij} -1\,.
\ee
In a  convenient choice of basis they are given by
\be
\vec a_1=(1,1,1)\,,\quad \vec a_2= (1,-1,-1)\,,\quad
\vec a_3=(-1,1,-1)\,,\quad \vec a_4=(-1,-1,1)\,.
\ee
The extreme static black-hole solutions are given by
\bea
ds^2 &=& - H^{-1/2}\, dt^2 + H^{1/2}\, (dr^2 + r^2 d\Omega^2)\,,\qquad
e^{-\ft12 \vec a_i\cdot\vec\varphi}= H_i\, H^{-1/4}\,,\\
A^i &=& (1-H_i^{-1})\, dt\,,\nn
\eea
where
\be
H= \prod_{i=1}^4 H_i\,,\qquad H_1= 1 + \fft{q_i}{r}\,.
\ee
The charge parameters $q_i$ should all be non-negative in order to
have a black-hole solution with no naked singularity, and an event horizon
at $r=0$.

   We can define Eddington-Finkelstein coordinates $u$ and $v$ in the
standard way:
\be
u=t-r_*\,,\qquad v=t+ r_*\,,\qquad r_*\equiv \int^r \sqrt{H(r')}\, dr'\,,
\label{uvr*}
\ee
in terms of which the static black-hole metrics become
\bea
ds^2 &=& -H^{-1/2}\, du^2 - 2 du dr + H^{1/2}\, r^2 d\Omega^2\,,\nn\\
&=& -H^{-1/2}\, dv^2 + 2 dv dr + H^{1/2}\, r^2 d\Omega^2\,.\label{EFmets}
\eea
The explicit expression for $r_*$ is complicated in general.

   The scalar wave equation $\square\psi=0$ becomes
\bea
0&=& -2\del_r\del_u\psi - 
   \fft1{r^2\, H^{1/2}}\, \del_r\big(r^2\, H^{1/2}\big)\, 
  \del_u\psi  +  \fft{1}{r^2\, H^{1/2}} \del_r(r^2\del_r\psi) +
  r^2 H^{-1/2} \Delta \psi\,,\nn\\
0&=& 2\del_r\del_v\psi + 
   \fft1{r^2\, H^{1/2}}\, \del_r\big(r^2\, H^{1/2}\big)\,
  \del_v\psi +  \fft{1}{r^2\, H^{1/2}} \del_r(r^2\del_r\psi) +
 r^2 H^{-1/2} \Delta \psi\,,\label{scalarEF}
\eea
where $\Delta$ is the scalar Laplacian on the unit 2-sphere.

\subsection{Inversion and conformal symmetry}
  
  If we define a new radial coordinate
\be
\tilde r= \fft{Q^2}{r}\,,\qquad \hbox{where}\ \ Q^4=\prod_i q_i\,,
\label{inversion}
\ee
it is easy to see that the metric becomes
\be
ds^2= \fft{Q^2}{\tilde r^2}\, d\tilde s^2\,,
\ee
where
\bea
d\tilde s^2 &=& -\widetilde H ^{-1/2}\, dt^2 + \widetilde H^{1/2}\,
(d\tilde r^2 + \tilde r^2\, d\Omega^2)\,,\\
\widetilde H &=& \prod_i \widetilde H_i \,,\qquad \hbox{where}\ \ 
  \widetilde H_i = 1 +\fft{\tilde q_i}{\tilde r}\,,\qquad
\tilde q_i = \fft{Q^2}{q_i}\,.\nn
\eea
Thus the metric written in terms of the $\tilde r$ radial coordinate
is conformally related to a metric of the original form, but with
redefined charge parameters. 

\subsection{The pairwise-equal charge specialisation}

   Because the charge parameters in the metric $d\tilde s^2$ are
different from those of the original metric $ds^2$, the inversion
transformation (\ref{inversion}) is not a conformal symmetry.  However,
it does become a conformal symmetry in the special case where the
four electric charges are set pairwise equal, without loss of generality,
\be
q_3=q_1\,,\qquad q_4=q_2\,,\label{pairwise}
\ee
since then we will have
\be
Q^2= q_1\, q_2\,,\qquad \tilde q_1= q_2\,,\qquad \tilde q_2= q_1\,,
\ee
and so the metric function $\widetilde H(\tilde r)$ is equal to the
original metric function $H(r)$.  Note also that in the
pairwise equal case the metric takes the very simple form, with
no irrational functions,
\be
ds^2_{\rm pwe} = -\Big[\Big(1+\fft{q_1}{r}\Big)
   \Big(1+\fft{q_2}{r}\Big)\Big]^{-1}\, dt^2 + 
  \Big(1+\fft{q_1}{r}\Big)
   \Big(1+\fft{q_2}{r}\Big)\, (dr^2+ r^2 d\Omega^2)\,.\label{pwmetric}
\ee
Note that the ERN solution arises as the
further specialisation where we set $q_1=q_2=q$ in (\ref{pwmetric})
(after sending $r\rightarrow r-q$ to express the metric in the usual
form).

   The explicit expression for the $r_*$ coordinate defined in (\ref{uvr*}),
which is rather complicated in the general 4-charge case, becomes very
simple in the case of pairwise-equal charges:
\be
r_* = r + (q_1+q_2)\, \log r - \fft{q_1 q_2}{r}\,.
\ee
The scalar wave equation, written in terms of advanced or retarded null
coordinates as in \eqref{scalarEF}, becomes quite simple also.  In 
particular, if we expand in a power series in $1/r^n$ around
infinity, or in $r^n$ around the horizon,
\be
\psi(r,u,\theta,\varphi)= \sum_{n\ge 1} \psi_n(u,\theta,\varphi)\, r^{-n}\,,
\qquad
\widetilde \psi(r,v,\theta,\varphi)= 
\sum_{n\ge0} \widetilde\psi_n(v,\theta,\varphi)\, r^n\,,
\ee
then there will be just a fixed finite number of terms in the various NP or
Aretakis charges.  By contrast, in the general 4-charge solutions
the number of terms will grow with $n$.

\subsection{General static extreme STU black holes}\label{dyonicsec}

   We saw above that the static extreme STU black holes with four
independent electric charges admit an inversion relation under which
a given black hole maps into a conformally-transformed member of the 
same family with different
values for the four charges.  The charges of the original and the transformed
black holes become the same in the special case where the four charges
are taken to be pairwise equal.

The most general static black holes in STU supergravity have eight independent
charges, with each of the four field strengths carrying both an electric
and a magnetic charge.  It is of interest to investigate whether an 
inversion relation of the kind we found for the extreme four-charge black
holes extends to the most general eight-charge extreme black holes.  We find
that the answer is no.  The eight-charge solutions are actually quite 
complicated (see, for example, \cite{chowcomp}), but for our purposes 
it will suffice to consider the simpler case where only one of the
four field strengths is turned on, and it carries independent electric
and magnetic charges.  For the purposes of describing this dyonic 
solution one can consistently truncate the STU theory to the so-called
``Kaluza-Klein theory,'' that results from dimensionally reducing pure 
five-dimensional gravity to four dimensions.

  The details of the extreme static Kaluza-Klein dyonic black hole are 
presented in appendix A, where we show that it does not map under inversion
into a conformally related extreme dyonic black hole, except for special
choices for the two charge parameters which then reduce to cases
encompassed by our previous discussion.  Of course, having established this
result for the truncated system of the Kaluza-Klein solutions, it follows
that there is also no inversion transformation relation generically for the
eight-charge black holes.

   Although our findings for the dyonic black holes are negative, the 
result does serve to highlight the fact that there is something rather
special about the case where each field strength in the STU theory
carries only a purely  electric or a purely magnetic charge, as in the 4-charge 
solutions we discussed previously.

\section{Correspondence between Aretakis and NP charges} \label{sec:corres}

In order to understand the existence of Aretakis charges on the horizon of 
general extreme black holes \cite{Aretakis:extremal}, we relate, via a 
conformal map, their geometry close to the horizon to the class of solutions 
that we introduced in section \ref{sec:NP}, namely \emph{weakly asymptotically 
flat} space-times.  Such space-times include asymptotically flat space-times 
as a subset.  

The existence of conserved charges near null infinity is not so surprising 
given that we have asymptotic Killing isometries that give rise to conserved 
charges \cite{WZ}, such as those identified by Newman and Penrose 
\cite{NP}.  If these NP charges are generated by an (asymptotically) 
conformal field, then they map on to the Aretakis charges on the extremal 
horizon via the conformal map that relates the two asymptotic space-times.  
We shall demonstrate this explicitly at the end of this section.  However, 
before that we derive the conformal map.

As was explained in section \ref{sec:A}, given an extreme black hole 
solution, we can always choose Gaussian null coordinates in an open patch 
in the neighbourhood of the horizon such that the metric takes the form 
given in \eqref{NHC}, which we repeat here for convenience
\begin{equation} \label{GNC}
 ds^2 = L(x)^2 \Big[- r^2 F(r,x) dv^2 + 2 dv dr \Big]  + 
\gamma_{IJ}(r,x) \Big(dx^I - r\, h^I(r,x) dv \Big) \Big(dx^J - 
r\, h^J(r,x) dv\Big).
\end{equation}
We have Killing vectors
\begin{equation}
 k=\partial/\partial v, \qquad m=\partial/\partial \phi \equiv 
\partial/\partial x^1.
\end{equation}
Moreover, as before we assume that\footnote{Apart from the leading order 
terms, we do not need to assume any analyticity properties for the metric 
functions.  This seems like a reasonable assumption.  For example, at null 
infinity, it is known that non-analytic terms may contribute at lower 
orders \cite{NP61,Damour:1985cm,christ}.}
\begin{equation} \label{GNC:falloff}
 F(r,x) = 1+\mathcal{O}(r),\quad  h^{\phi} = \mathcal{O}(1), 
\quad h^2 = \mathcal{O}(r), \quad \gamma_{IJ} = 
\tilde{\gamma}_{IJ}(x) + \mathcal{O}(r),
\end{equation}
where $\tilde{\gamma}_{IJ}$ is the metric on a compact space that 
is topologically spherical \cite{hawking}.  Now, we consider the
coordinate change
\begin{equation}
  v \rightarrow u, \qquad r \rightarrow \frac{1}{r},
\end{equation}
which gives
\begin{equation} \label{conWAF}
 ds^2 = \frac{L^2}{r^2} \Big\{ - F du^2 - 2 du dr + r^2 h_{IJ} (dx^I - C^I du) (dx^J - C^J du) \Big\},
\end{equation}
where
\begin{equation}
 C^I = h^I/r, \qquad h_{IJ} = L^{-2} \gamma_{IJ}.
\end{equation}
Using the fall-off conditions in \eqref{GNC:falloff}
\begin{equation} \label{WAF:falloff}
 F(r,x) = 1 + \mathcal{O}(1/r), \quad C^\phi = \mathcal{O}(1/r), \quad C^{2} = \mathcal{O}(1/r^2),\quad  h_{IJ} = \omega_{IJ} + \mathcal{O}(1/r)
\end{equation}
with $\omega_{IJ}$ the metric on a topological sphere.  We identify the conformally related metric in \eqref{conWAF} as that of a weakly asymptotically flat space-time, with $\beta = 0$, 
as defined in section \eqref{sec:NP}.

In summary, we have found that the metric in the neighbourhood of an extremal horizon is conformally related, see equation \eqref{conWAF}, to that of a weakly asymptotically flat metric.   Thus, the Aretakis charge on the horizon generated by a conformal field maps on to the NP charge generated by that field at the future null infinity of the weakly asymptotically flat space-time 
associated with the extreme black hole via this conformal correspondence.

\subsection{Massless scalars} \label{sec:scalar}

More concretely, we take as an example the massless scalar field that was 
considered in section \ref{sec:charge},
\begin{equation} \label{KGeqn}
 \Box_g \psi = 0.
\end{equation}
Assuming that the background extreme black hole, with metric  $g$, is 
Ricci scalar-flat~\footnote{Any solution of the Einstein equation with $\Lambda = 0$ and traceless energy-momentum tensor $T_{a}{}^a = 0$ will have vanishing Ricci scalar.  For backgrounds with non-vanishing Ricci scalar, one would naturally consider the conformally covariant wave operator.}, the wave operator $\Box_g$ coincides with the conformally 
covariant wave operator
\begin{equation}
 \mathcal{O}_g = \Box_g - \ft16 R(g).
\end{equation}
In the previous discussion, we have identified a self-inverse diffeomorphism
\begin{equation}
 \iota: (v, r, x^I) \longrightarrow \left(u=v,\, r'=\frac{1}{r},\, 
x^I\right), \qquad \iota^{-1}: (u, r, x^I) \longrightarrow 
\left(v=u,\, r'= \frac{1}{r},\, x^I\right),
\end{equation}
such that given the metric $g$ in Gaussian null coordinates obeys
\begin{equation}
 (\iota^{-1})_*(g) = \Omega^2 \, \tilde{g},
\end{equation}
where
\begin{equation}
 \Omega = L/r
\end{equation}
and $\tilde{g}$ is the associated weakly asymptotically flat metric.

Now, suppose we have a solution of the wave equation on the associated 
background:
\begin{equation}
 \Box_{\tilde{g}}\, \tilde{\psi} = 0.
\end{equation}
Although the Ricci scalar $R(\tilde{g})$ will not generally be zero, the 
solution is nevertheless asymptotically Ricci-flat.  Hence for large $r$, 
which is the region in which we are interested, we have
\begin{equation}
 \mathcal{O}_{\tilde{g}} \tilde{\psi} \approx 0,
\end{equation}
i.e.~$\mathcal{O}_{\tilde{g}} \tilde{\psi}$ is asymptotically zero.  Given 
such a solution $\tilde{\psi}$, the conformal covariance of the
operator $\mathcal{O}$ implies that
\begin{equation}
 \mathcal{O}_{\Omega^2 \tilde{g}}\, \psi = 
\Box_{\Omega^2 \tilde{g}} \, \psi = 0,
\end{equation}
where
\begin{equation}
 \psi = \Omega^{-1} \, \tilde{\psi}
\end{equation}
and we have used the fact that $\Omega^2 \tilde{g}$ is Ricci scalar-flat.  In 
conclusion, we have generated, using this map, a solution of the wave 
equation on the horizon of the original solution of interest,
\begin{equation} \label{scalarmap}
 \psi(v,r,x^I) = \frac{L}{r}\, \tilde{\psi}\left(v, \frac{1}{r},x^I \right).
\end{equation}

This relation can then be used to map the NP charge at null infinity 
of the dual space-time to the Aretakis charge at the extremal horizon.

\subsection{Example: Extreme Reissner-Nordstr\"om solution} \label{sec:exERN}

We demonstrate this correspondence explicitly for the ERN solution, without recourse to the special conformal isometry \cite{Couch1984} that it exhibits.  Starting with coordinates adapted to the extremal horizon, the metric takes the form \eqref{ERN:GNC},
\begin{equation} \label{ERN:GNC1}
 ds^2 = M^2 \left[ - \frac{r^2}{(1+Mr)^2} dv^2 + 2 dv dr \right] + M^2 (1+Mr)^2 d \Omega^2.
\end{equation}
Now we let
\begin{equation}
 v \rightarrow u, \qquad r \rightarrow \frac{1}{r},
\end{equation}
which gives
\begin{equation}
 ds^2 = \frac{M^2}{r^2} \left\{ - \left(1+\frac{M}{r}\right)^{-2} du^2 - 2 du dr + r^2 \left(1+\frac{M}{r}\right)^{2} d \Omega^2 \right\}.
\end{equation}
Hence, we identify the weakly asymptotically flat space-time dual to the ERN metric to be
\begin{equation} \label{dualERN}
 ds^2 = - \left(1+\frac{M}{r}\right)^{-2} du^2 - 2 du dr + r^2 \left(1+\frac{M}{r}\right)^{2} d \Omega^2.
\end{equation}
The determinant of this metric is given by
\begin{equation}
 \sqrt{-g} = r^2 \left(1+\frac{M}{r}\right)^{2} \sin \theta,
\end{equation}
from which we read off (compare with \eqref{det:asymp})
\begin{equation}
 \zeta = \left(1+\frac{M}{r}\right)^{2}.
\end{equation}
Substituting this expression into equation \eqref{NPcharge} gives
\begin{equation}
 H_{NP} = - 2\lim_{r\to\infty} \int d\omega \  \left[  r^2 \partial_r (r \tilde{\psi}) - M r\, \tilde{\psi} \right],
\end{equation}
where $\tilde{\psi}$ solves the wave equation on the associated weakly 
asymptotically flat background given by metric \eqref{dualERN}.  Assuming for 
convenience that for large $r$
\begin{equation} \label{psitexp}
 \tilde{\psi} = \frac{\tilde{\psi}^{(1)}}{r} + \frac{\tilde{\psi}^{(2)}}{r^2} + \ldots,
\end{equation}
we find that
\begin{equation}
 H_{NP} = 2\int d\omega\ \left(\tilde{\psi}^{(2)}+M \tilde{\psi}^{(1)} \right).
\end{equation}
Consider a solution $\psi$ on the original ERN background, given by the
metric \eqref{ERN:GNC1}, which we assume to take the form
\begin{equation} \label{psiexp}
 \psi = \psi^{(0)} + r \psi^{(1)} + \ldots
\end{equation}
near the horizon at $r=0.$  Equation \eqref{scalarmap} gives that
\begin{equation}
 \psi(v,r,x^I) = \frac{M}{r} \tilde{\psi}(v,\fft1{r},x^I).
\end{equation}
Substituting the expansions of the scalar fields \eqref{psitexp} and 
\eqref{psiexp} gives
\begin{equation}
 \psi^{(0)} = M \tilde{\psi}^{(1)}, \quad \psi^{(1)} = M \tilde{\psi}^{(2)}.
\end{equation}
Hence we find
\begin{equation}
 H_{NP} = \frac{2}{M} \int d\omega \
  \left({\psi}^{(1)}+M {\psi}^{(0)} \right).
\end{equation}
Comparing this expression with the Aretakis charge \eqref{AERN}, 
written in terms of the expansion of $\psi$ \eqref{psiexp}, we find that
\begin{equation}
 H_{Aretakis} = M^{3}\, H_{NP}.  
\end{equation}
Thus, up to an irrelevant factor, we have identified the NP charge at 
null infinity of the associated weakly asymptotically flat solution with 
the Aretakis charge on the horizon of the ERN black hole.

\section{Discussions} \label{sec:dis}

In this paper, we have found a correspondence between extreme black holes and 
weakly asymptotically flat space-times. These are space-times for which 
Bondi coordinates can be introduced.  However, the metric functions fall off 
at a slower rate than that which asymptotic flatness would demand.  
Furthermore, the spatial sections of ``future null infinity'' are only 
required to be compact, rather than specifically $S^2$.  
In section \ref{sec:NP}, we found that weakly asymptotically flat 
space-times admit an NP charge at null infinity.  What distinguishes 
asymptotically flat space-times in four dimensions is the existence of an 
enhanced symmetry group at null infinity---the BMS group.  The existence of 
NP charges for the aformentioned more general space-times suggests that 
there cannot be a direct relation between NP charges and the BMS group.  
Another indication of this is that NP charges arise in higher-dimensional
weakly asymptotically flat space-times, even though there is no 
BMS group in these cases.  

  Nevertheless, one might expect the BMS group to play a r\^ole in the 
existence of charges and the representations that they belong to.  A question 
that remains unanswered, as far as we are aware, is whether NP charges have 
special additional features in space-times where the asymptotic symmetry 
group is enhanced, and if so how?  We hope to return to this issue in 
the future.

Going back to the correspondence that we introduced in section 
\ref{sec:corres}, we demonstrated it explicitly for the ERN black hole in 
section \ref{sec:exERN}.  Of course, the existence of the Aretakis charges in 
terms of NP charges for the ERN black hole had already been explained in the 
literature.  Surely, a more interesting example to consider would be the 
extreme Kerr(-Newman) black hole.  The first step in this construction 
would be to construct Gaussian null coordinates in the neighbourhood of 
its horizon, such that the metric takes the form \eqref{NHC}.  Given how 
non-trivial it is to construct Bondi coordinates for the Kerr metric 
\cite{PI, bishop, fletcher}, we do not expect this to be an easy task.  
Nevertheless, it may be an illuminating one.

\section*{Acknowledgements}

We would like to thank Stefanos Aretakis for useful discussions.  
H.G.\ and M.G.\ would like to thank the Mitchell Institute for Fundamental 
Physics and Astronomy, Texas A\&M University, where this work was initiated, and C.N.P.\ in particular, 
for hospitality. M.G.\ would like to thank the 
Max-Planck-Institut f\"ur Gravitationsphysik (Albert-Einstein-Insitut), Potsdam, 
for hospitality during the course of this work. M.G.\ is partially 
supported by grant no.\ 615203 from the European Research Council under the FP7.
C.N.P.\ is partially supported by DOE grant DE-FG02-13ER42020.

\appendix

\section{Inversion and the extreme dyonic Kaluza-Klein black holes} \label{app:dyonic}

   Here we present some details of the extreme static dyonic black hole
in the truncation of STU supergravity to the Kaluza-Klein theory that
results from the dimensional reduction of five-dimensional pure
gravity on a circle.  The four-dimensional theory is described by the
Lagrangian
\be
{\cal L}= \sqrt{-g}\, \Big[ R - \ft12 (\del\varphi)^2 -
           \ft14 e^{-\sqrt3 \varphi}\, F^2\Big].
\ee
The extreme static dyonic black hole is given by (see, for example, 
\cite{lupapo} for the details in the conventions we are using here)
\bea
ds^2 &=& -(H_1 H_2)^{-1/2}\, dt^2 + (H_1 H_2)^{1/2}\, \Big( dr^2 +
   r^2\, d\Omega^2\Big)\,,\nn\\
\varphi &=& \fft{\sqrt3}{2}\, \log \fft{H_2}{H_1}\,,\nn\\
H_1 &=& 1 + \fft{4 Q^{2/3}\, \sqrt{P^{2/3} + Q^{2/3}}}{r} +
  \fft{8 P^{2/3}\, Q^{4/3}}{r^2}\,,\nn\\
H_2 &=& 1 + \fft{4 P^{2/3}\, \sqrt{P^{2/3} + Q^{2/3}}}{r} +
  \fft{8 Q^{2/3}\, P^{4/3}}{r^2}\,,\label{dyonmet}
\eea
where $P$ and $Q$ are the magnetic and electric charges.  The gauge
potential can be found also in \cite{lupapo}, but we shall not need it here.

   We now consider the inversion transformation under which we define a 
new radial coordinate
\be
\tilde r= \fft{8P Q}{r}\,.
\ee
We then find that the metric (\ref{dyonmet}) can be written as
\be
 ds^2 = \fft{8 P Q}{\tilde r^2}\, d\tilde s^2\,,
\ee
where
\be
d\tilde s^2 = -(\widetilde H_1 \widetilde H_2)^{-1/2}\, dt^2 + 
(\widetilde H_1 \widetilde H_2)^{1/2}\, \Big( d\tilde r^2 + 
                \tilde r^2\, d\Omega^2\Big)\,,
\ee
with
\bea
\widetilde H_1 &=& 1 + \fft{4 (PQ)^{1/3}\,\sqrt{P^{2/3} + Q^{2/3}}}{\tilde r} +
  \fft{8 P^{4/3}\, Q^{2/3}}{\tilde r^2}\,,\nn\\
\widetilde H_2 &=& 1 + \fft{4 (PQ)^{1/3}\, \sqrt{P^{2/3} + Q^{2/3}}}{\tilde r} +
  \fft{8 Q^{4/3}\, P^{2/3}}{\tilde r^2}\,.
\eea

We see that for general $P$ and $Q$ there is no way to map
the $H_i$ functions into the $\widetilde H_i$ functions by choosing different
parameter values.  It can be done only if $P=0$ or $Q=0$ or $P=Q$.  
The first of
these is just a single electric charge specialisation of the previous 
4-charge
case.  The second is an equivalent single magnetic charge case.  
The third,
although dyonic, is merely a duality rotation of the ERN black hole.

\bibliographystyle{utphys}
\bibliography{nparetakis}

\end{document}